\documentclass[useAMS,usenatbib]{mn2e}
\usepackage{graphicx,times}

\newcommand{\dm}{{\rm d}}

\title[The effects of feedback on the morphology of galaxies]{
The effects of feedback on the morphology of galaxy discs}
\author[T. Okamoto et al.]{Takashi Okamoto$^{1}$$^{,2}$\thanks{E-mail:
takashi.okamoto@durham.ac.uk}, Vincent R. Eke$^{1}$, Carlos
S. Frenk$^{1}$ and Adrian Jenkins$^{1}$\\
$^{1}$Institute for Computational Cosmology, Department of Physics,
University of Durham, South Road, Durham DH1 3LE, UK\\ 
$^{2}$Department of Theoretical Astronomy, National Astronomical
Observatory Japan, Osawa, Mitaka, Tokyo 181-8588 Japan} 
\begin{document}

\date{}

\pagerange{\pageref{firstpage}--\pageref{lastpage}} \pubyear{2004}

\maketitle
 
\label{firstpage}

\begin{abstract}
We have performed hydrodynamic simulations of galaxy formation in a
cold dark matter ($\Lambda$CDM) universe. We have followed galaxy
formation in a dark matter halo, chosen to have a relatively quiet
recent merger history, using different models for star formation and
feedback. In all cases, we have adopted a multi-phase description of
the interstellar medium and modelled star formation in quiescent and
burst modes. We have explored two triggers for starbursts: strong
shocks and high gas density, allowing for the possibility that stars
in the burst may form with a top-heavy initial mass function. We find
that the final morphology of the galaxy is extremely sensitive to the
modelling of star formation and feedback. Starting from identical
initial conditions, galaxies spanning the entire range of Hubble
types, with $B$-band disc-to-total luminosity ratios ranging from 0.2
to 0.9, can form in the same dark matter halo. Models in which
starbursts are induced by high gas density (qualitatively similar to
models in which feedback is produced by AGN) generate energetic winds
and result in galaxies with an early-type morphology. Models in which
the starbursts are induced by strong shocks lead to extended discs. In
this case, the feedback associated with the bursts suppresses the
collapse of baryons in small haloes, helping to create a reservoir of
hot gas that is available for cooling after $z\simeq 1$, following the
bulk of the dynamical activity that builds up the halo. This
gas then cools to form an extended, young stellar disc.
\end{abstract}


\section{INTRODUCTION}

Understanding galaxy formation is a challenging problem whose solution
will require a concerted approach combining observational,
semi-analytic and numerical work. There have been substantial advances
on all these fronts in the past decade, but fundamental questions
remain unanswered. One of the most troublesome is the inability to
produce realistic spiral galaxies in numerical simulations that start
from initial conditions appropriate to the concordance cold dark matter
cosmology ($\Lambda$CDM) and assume prescriptions for gas and stellar
processes analogous to those included in semi-analytic models, or
inferred directly from observations.

Pioneering N-body/gasdynamical simulations by \citet{kg91},
\citet{nb91} and \citet{kat92}, already included many of the processes, 
which are generally regarded as essential for galaxy formation:
radiative cooling, star formation, and feedback effects generated by
energy released from associated with supernovae (SNe) and stellar
winds. The simulations by \citet{nb91}, and later by
\citet{nw94} and \citet{nfw95}, assumed CDM initial conditions. 
They uncovered a fundamental problem that prevents the formation of
disc-dominated galaxies: the so-called ``angular momentum
problem.'' This is a generic problem that can be traced back to very
nature of the hierarchical clustering process characteristic of
structure growth from CDM initial conditions \citep{fwed85}. At early
times, small, dense cold dark matter haloes form. Radiative cooling is
very efficient and a large fraction of their gas, some of which
will turn into stars, cools into their centres. As haloes merge to form
larger objects, their orbital angular momentum is drained by
dynamical friction and exported to the dark matter at the outskirts of
the new haloes. Much of the original angular momentum of the baryonic
material is lost and the resulting galaxies become too centrally 
concentrated.  

\citet{wee98} and \citet{eew00} showed that if cooling is
artificially suppressed until the host haloes are well established,
then the simulations can produce galaxies that are less centrally concentrated 
and have higher specific angular momenta.  
Two ways to prevent the
early collapse of protogalactic clouds have been proposed: stronger
feedback than that provided by standard treatments of supernovae
\citep{sgv99, tc01} and a modification of the cosmological
framework, replacing CDM with warm dark matter (which does not induce
small-scale fluctuations; \citet{sd01} and \citet{gov04}). There are
also indications that spurious numerical effects, arising from the very
nature of the smoothed particle hydrodynamics (SPH) technique used in
the simulations, cause transfer of angular momentum from the cold gas
disc to the hot halo gas, significantly contributing to the angular
momentum problem \citep{oka03}. Aside from numerical effects, it is
clear that the angular momentum problem in simulations is telling us
something quite fundamental about the nature of star formation and
feedback processes in galaxy formation.

Some recent simulations have yielded more promising disc galaxies
in the CDM framework. \citet{sgp03} were able to generate a variety of
morphological types, including discs, by assuming a self-propagating
star formation model combined with very efficient SNe feedback. In
their model, star formation proceeds in two modes, ``early time'' and
``late time''. In the early time mode, star formation is very
efficient and the feedback is very strong; in the late time mode, the
star formation efficiency is low and there is no feedback.
\citet{gov04} produced a disc galaxy employing standard star formation and 
feedback recipes and claimed that numerical resolution is the primary
cause of the angular momentum problem. Abadi et al. (2003a, 2003b)
were able to generate a galaxy resembling early-type spirals for which
they calculated detailed photometric and dynamical properties.
\citet{rob04} adopted the multi-phase model for the star-forming interstellar 
medium (ISM) of \citet{sh03}, which stabilises gaseous discs against
Toomre's instability, and produced a galaxy having an exponential
surface brightness profile. 
Note that earlier work by \citet{sn99} followed galaxy formation in randomly 
chosen haloes, while \citet{aba03a} and  \citet{gov04} selected 
haloes that have no major mergers at low redshift ($z < 2$ in Governato et al.).  
In the $\Lambda$CDM model, haloes with such quiet merger histories, 
while favourable for disc formation, have a number density today which is too low 
to account for the observed number density of spiral galaxies.  

While early simulations already included simplified treatments of
chemical evolution \citep{sm95, rvn96, ber99}, more recent simulations
have considered Type Ia supernovae (SNe Ia) in addition to Type II
supernovae (SNe II), sometimes relaxing the instantaneous recycling
approximation (IRA) \citep{lpc02, kg03, kob04}. In particular,
\citet{kg03} have argued that the non-instantaneous nature of SNe~II and
SNe~Ia is important for the dynamical evolution of galaxies.
\citet{bro04} have shown that the chemical abundance of the halo stars 
could be an additional constraint for the modelling of feedback. 

In this paper, we present a new series of simulations of galaxy
formation. The main difference with previous work is that we consider
an unconventional star formation model which, however, has been claimed
to be {\it required} to explain the properties of high redshift
sub-millimeter and Lyman-break galaxies \citep{bau04}, as well as the
metallicity of the intracluster medium \citep{nag04} and of elliptical
galaxies \citep{nag05}. In this model, star formation normally
proceeds in a quiescent mode with a standard initial mass function
(IMF). However, when a major merger occurs, it triggers a burst of star 
formation with a top-heavy IMF.  The distinction between
quiescent and burst modes is not particularly controversial but the
adoption of a top-heavy IMF is. \citet{bau04} claimed that the number
density of sub-millimeter galaxies, in particular, is impossible to
explain without a top-heavy IMF in bursts,  while
\citet{nag04} claimed that the ratio of alpha to iron peak elements in the
intracluster medium can only be understood if bursts have a top-heavy
IMF. A review of the observational and theoretical evidence for and
against a top-heavy IMF may be found in these papers and references
therein.

Our main result is that the adoption of a top-heavy IMF in bursts
eases the formation of a large disc primarily because at early times,
when mergers are more important, more energy per unit of mass turned
into stars is made available for feedback. However, we also
find that varying the criteria for a burst or varying the IMF in
bursts can have a strong effect on the final morphology of the
simulated galaxy. 

This paper is organised as follows. We describe our simulation code
and model for star formation and feedback in Section~2. We present the
details of our simulations in Section 3 and the results in Section
4. Finally, in Section~5, we present a discussion and summary of our
conclusions.

\section[]{SIMULATION CODE}

We use the parallel PM-TreeSPH code GADGET2 \citep{spr05}, a successor 
of the TreeSPH code GADGET \citep{syw01}. The hydrodynamics are solved using
an SPH algorithm \citep{luc77, gm77} and a ``conservative entropy''
formulation that manifestly conserves energy and entropy
\citep{sh02}.

Here, we briefly describe modifications we have made to the code so as
to adapt it for our purposes.  In the conservative entropy
formulation, the smoothing length of an SPH particle should be given
by \begin{equation} \frac{4 \pi}{3}h_i^3 \rho_i = M_{\rm ngb} =
const., \label{nngb} \end{equation} where $h_i$ and $\rho_i$ are the
smoothing length and density of the $i$th SPH particle respectively,
and $M_{\rm ngb}$ represents the mass in its smoothing volume.  We set
$M_{\rm ngb} = 40 m_{\rm orig}$ throughout this paper, where $m_{\rm
orig}$ is the original SPH particle mass.  The default implementation
of GADGET2 keeps the number of neighbouring particles in the smoothing
volume constant, whereas we chose to keep the mass resolution constant
as in equation (\ref{nngb}).  Since in our simulations the gas
particle masses vary due to star formation and feedback, these two
choices are not equivalent.  We also adopt the phase decoupling
technique introduced by \cite{oka03} to avoid spurious angular
momentum transfer from cold gas discs to surrounding hot halo gas.  We
will describe the modelling of cooling, star formation, and feedback
in the following subsections.

\subsection{Gas cooling}

We calculate the cooling/heating rate and ionization state of each
particle by assuming collisional ionization equilibrium and the
presence of an evolving but uniform UV background \citep{hm96} that is
switched on at $z = 6$. Inverse Compton cooling is also included. In
order to take the metallicity dependence into account, we use the
appropriate cooling tables given by \citet{sd93} at $T > 10^4$ K. The
coolest gas in overdense regions typically has a temperature $T \simeq
10^4$ K, because we do not include molecular cooling or metal cooling
below $10^4$ K.

Metals are carried by particles and once assigned to a particle, they 
remain with it. Nonetheless, effective mixing
takes place because we use the smoothed metallicity (smoothed in the same
way as other SPH quantities) when computing the cooling rates.  

\subsection{Star formation and feedback}

In general, star formation in numerical simulations is modelled as 
\begin{equation}
\frac{d\rho_*}{dt} = c_* \frac{\rho_{\rm g}}{t_{\rm dyn}}, 
\label{schmidt}
\end{equation}
where $\rho_*$, $\rho_{\rm g}$, $t_{\rm dyn} = (4 \pi G \rho_{\rm g})^{-1/2}$, 
and $c_*$  are the stellar density, the gas density, the local dynamical time, 
and a dimensionless star formation efficiency parameter respectively.

Typically, $c_* \simeq 1/30$ has been used in simulations of disc 
formation \citep[e.g.][]{sn99, tc01, aba03a, gov04, rob04} 
in order to reproduce the observed gas mass fraction and/or the Kennicutt 
(1998) law. 
\citet{mez03} showed that the same star formation model used by \citet{aba03a} 
could also lead to the formation of an elliptical galaxy if the initial 
conditions were such that a major merger took place at low redshift. 
On the other hand, some simulations of elliptical galaxy formation,   
have used large values of $c_* = 0.5 \sim 1$ \citep[e.g.][]{kg04, kob04},
often with strong feedback, in order to reproduce the observed sizes 
and/or colours of elliptical galaxies. 
Sommer-Larsen et al. (2003) combined these two star
formation recipes by introducing `early' and `late' star formation
modes.  In the early star formation mode, they adopted $c_* = 1$,
together with a low threshold density for star formation and very
strong feedback while, in the late star formation mode, they adopted
$c_* = 1/40$, and prevented any feedback. With this prescription, they
were able to reproduce a wide range of morphological types in their
simulations.  

We consider two physically motivated star formation modes, which we
call `quiescent' and `burst'. We model the quiescent mode as
self-regulated star formation using a multi-phase ISM model based on
that developed by \cite{sh03}, but modified to avoid the instantaneous
recycling approximation (IRA) in which short-lived stars assumed to die 
immediately as they form.  
We use the same model for the burst mode, except that we adopt a shorter 
star formation timescale ($c_* = 0.5$) and a flatter IMF with the intention 
of making self-regulation an unstable process. The motivation for these 
choices was discussed in \S~1.

\subsubsection{A multi-phase model for star forming gas}

Following \citet{sh03}, the dense ISM gas is pictured as a two-phase
fluid consisting of cold clouds and an ambient hot phase whose energy
is supplied by SNe explosions.  We now briefly explain our model,
including the modifications we have implemented.

In \citet{sh03}, the densities of cold clouds 
($\rho_{\rm c}$) and the hot phase ($\rho_{\rm h}$) are related to the 
total gas density ($\rho$) as $\rho = \rho_{\rm c} + \rho_{\rm h}$. 
By considering the volume that each phase occupies, we impose the following
relations: 
\begin{eqnarray}
\label{dens}
\frac{M_{\rm SPH}}{\rho} &=& \frac{M_{\rm c}}{\rho_{\rm c}} + 
\frac{M_{\rm h}}{\rho_{\rm h}}, \\
\label{mass}
M_{\rm SPH} &=& M_{\rm c} + M_{\rm h}, 
\end{eqnarray}
where $M_{\rm SPH}$, $M_{\rm c}$, and $M_{\rm h}$ are the particle 
mass, mass in cold clouds, and mass in the hot phase 
associated with the particle.
We further assume the cold clouds are in pressure equilibrium with the hot 
ambient phase, namely
\begin{equation}
\rho u_{\rm eff} = \rho_{\rm c}u_{\rm c} = \rho_{\rm h}u_{\rm h}, 
\label{pequil}
\end{equation}
where $u_{\rm eff}$, is the effective specific internal energy of 
an SPH particle, and $u_{\rm c}$ and $u_{\rm h}$ are the specific 
internal energies of cold clouds and hot phase.

We describe our multi-phase model, first assuming the IRA, 
in which star formation, cloud formation by thermal instability, 
evaporation of clouds, and heating of the hot phase by SN explosions 
are included.  This model is almost identical
to that of \citet{sh03} except for the definition of the density of
each phase and the metallicity dependence.  We will then show how to
relax the IRA.

Star formation takes place on a star formation timescale $t_*$: 
\begin{equation}
\frac{\dm M_*}{\dm t} = \frac{M_{\rm c}}{t_*}, 
\label{SFR}
\end{equation}
where $M_*$ is the stellar mass. 
The star formation timescale, $t_*$, is taken to be compatible with equation 
(\ref{schmidt}) - for $\rho>\rho_{\rm th}$ we have:
\begin{equation}
t_*(\rho) = t_*^0 \left(\frac{\rho}{\rho_{\rm th}}\right)^{-1/2}, 
\end{equation}
where the value of $t_*^0$ is chosen to match the Kennicutt law 
\citep{ken98} and $\rho_{\rm th}$ is a threshold density above which 
SPH particles become multi-phase and are eligible to form stars. For
$\rho<\rho_{\rm th}$, no star formation takes place.

The short-lived stars immediately die and return mass and release
energy, $\epsilon_{\rm SN}$, for each solar mass in stars formed, as
SNe II. The heating rate arising from SNe is therefore
\begin{equation}
\left.\frac{\dm}{\dm t}(M_{\rm h}u_{\rm h})\right|_{\rm SN}
= \epsilon_{\rm SN} \frac{\dm M_*}{\dm t}. 
\label{SN}
\end{equation}

We assume that SN explosions also evaporate the cold clouds, and
transfer gas back into the ambient hot phase.
\begin{equation}
\left.\frac{\dm M_{\rm c}}{\dm t}\right|_{\rm EV} 
= A \frac{\epsilon_{\rm SN}}{u_{\rm SN}} \frac{M_{\rm c}}{t_*}.  
\label{EV}
\end{equation}
Here, $A$ is the efficiency parameter  
and the specific supernova energy, $u_{\rm SN}= \epsilon_{\rm
SN}/\beta$, where  
$\beta$ is the mass fraction of short-lived stars. 
Following \citet{mo77}, the efficiency parameter has the density dependence 
\begin{equation}
A(\rho) = A_0 \left( \frac{\rho}{\rho_{\rm th}} \right)^{-4/5}. 
\label{evparam}
\end{equation}

Finally, we assume the cold clouds form and grow through thermal 
instability, that is 
\begin{equation}
\left. \frac{\dm M_{\rm c}}{\dm t} \right|_{\rm TI}
= - \left. \frac{\dm M_{\rm h}}{\dm t} \right|_{\rm TI}
= \frac{\Lambda_{\rm net}(\rho_{\rm h}, u_{\rm h}, Z)}{u_{\rm h} - u_{\rm c}} 
\frac{M_{\rm h}}{\rho_{\rm h}}, 
\label{COOL}
\end{equation}
where $\Lambda_{\rm net}$ is the cooling function for gas of
metallicity $Z$. We will assume constant temperature, $T_{\rm c}
\simeq 1000$ K, for cold clouds. We will use this instantaneous
recycling approximation to fix the model parameters $t_0^*$,
$\rho_{\rm th}$, $\epsilon_{\rm SN}$, and $A_0$ so as to reproduce the
Kennicutt law in subsection \ref{quiescent}.

\subsubsection{Removing the instantaneous recycling approximation} 

We do not use the IRA in our cosmological simulations.  
In this section, we describe how we model feedback and chemical evolution. 
We assume that each stellar particle represents
a single stellar population, and so star formation must be treated 
statistically.  Using equation (\ref{SFR}), a qualifying particle
spawns a new stellar particle of mass $m_* = m_{\rm orig}/N_{\rm g}$
during a timestep $\Delta t$ with probability:
\begin{equation} 
p = \frac{M_{\rm c}}{m_*} \left[1- \exp\left(-\frac{\Delta t}{t_*}\right)\right].
\label{SFprob} 
\end{equation} 
We use $N_{\rm g} = 3$ and have confirmed that our results do not
change for $N_{\rm g} \geq 2$.  When an SPH particle spawns a stellar
particle, the mass of cold clouds is reduced by $m_*$ and the
following new specific energy $u_{\rm eff}'$ and new density $\rho'$ 
(in the code we compute the effective entropy instead of $u_{\rm eff}$) 
are given to the particle assuming the particle volume $M_{\rm SPH}/\rho$ 
remains constant during the timestep: 
\[u_{\rm eff}' = \frac{M_{\rm SPH} u_{\rm eff} - m_* u_{\rm
c}}{M_{\rm SPH} - m_*},\]
\[\rho' = \rho \ \frac{M_{\rm SPH} - m_*}{M_{\rm SPH}}.\] 
We then update $M_{\rm SPH}$.  

Each stellar particle has its own age, metallicity, and IMF. 
We define the IMF as the number of stars per logarithmic interval 
of stellar mass per unit total mass of stars formed:
\begin{equation}
\phi(m) \equiv \dm N/\dm \ln m \propto m^{-x};   m_l < m < m_u, 
\label{IMF}
\end{equation}
where $m_l$ and $m_u$ are the lower and upper mass limits of the IMF.
The IMF has the following normalisation:
\begin{equation}
\int_{m_l}^{m_u} m \phi(m) \dm \ln m 
= \int_{m_l}^{m_u} \phi(m) \dm m = 1. 
\end{equation}

The recycled mass fraction from an evolving population of stars is 
a function of time. For a population of stars formed with an IMF 
$\phi(m)$ at time $t=0$, the mass fraction returned to the ISM
by time $t$ is 
\begin{equation}
E(\leq t) = \int_{M(t)}^{m_{\rm u}}[m-M_r(m)]\phi(m)\frac{\dm m}{m}, 
\end{equation}
where $M(t)$ is the initial mass of a star just reaching the end of its lifetime 
at $t$, and $M_r(m)$ is the mass of the remnant left by a star with 
initial mass $m$. The lifetimes are taken from \citet{pcb98} for 
massive stars and \citet{mar01} for intermediate and low mass stars. 

The cumulative number of SNe II explosions up to time $t$ (per solar
mass of stars formed), $R_{\rm II}(\le t)$, and the mass fraction of
metals of the $i$-th element, $E_i(\le t)$, are given by
\begin{eqnarray}
R_{\rm II}(\le t) 
&=& \int_{{\rm max}(M(t), 8M_\odot)}^{m_{\rm u}} \phi(m) \frac{\dm m}{m}, \\
E_i(\le t)  
&=& \int_{M(t)}^{m_{\rm u}} M_i(m) \phi(m) \frac{\dm m}{m}, 
\end{eqnarray}
where $M_i(m)$ is the mass of the $i$-th element released from 
a star with mass $m$. 
As $E_i$ includes metals present when the stars formed, 
the pre-existing metals must be subtracted in order to estimate 
chemical yields, $p_i(t)$. 
If the metallicity of stars with respect to the $i$-th element at 
their birth is $Z_i^0$, $E_i$ can be divided into two terms, 
\begin{equation}
E_i(\le t) 
= p_i(\le t) + Z_i^0 E(\le t), 
\end{equation}
where
\begin{equation}
p_i(\le t) 
= \int_{M(t)}^{m_{\rm u}} y_i(m)\phi(m)\frac{\dm m}{m}, 
\end{equation}
and $y_i(m)$ indicates the mass of newly produced $i$-th element in 
a star with mass $m$. 
Although $M_i$ and $y_i$ depend on the initial stellar metallicity 
in principle, we neglect this dependence and use the values for solar 
initial metallicity throughout, which are taken from Portinari et al. (1998). 

We consider energy and chemical feedback from SNe Ia as well. 
We calculate the number of SNe Ia explosions using the scheme 
of \citet{gr83}, with parameters updated according to 
Portinari et al. (1998). 
The progenitors of SNe Ia are assumed to be binary systems with 
initial masses in the range 
$m_{\rm B, low} < m_{\rm B} < m_{\rm B, up}$, 
where $m_{\rm B} \equiv m_1 + m_2$, and $m_1$ and $m_2$ 
are the initial masses of the primary and secondary, 
respectively. 
We assume $m_{\rm B, up} = 2 m_{\rm 1, max}$, where $m_{\rm 1, max}$
is the largest single-star mass for which the endpoint is a C-O 
white dwarf. 
The binary star systems that are the progenitors of SNe Ia are 
assumed to have an initial mass function $B\phi(m_{\rm B})$, where 
$\phi(m)$ is the same as for the single-star IMF. 
The distribution of mass fraction for the secondary, 
$\mu = m_2/m_{\rm B}$, is assumed to have the form 
(normalised over the range $0 < \mu < 1/2$)
\begin{equation}
f(\mu) = 2^{1+\gamma}(1+\gamma)\mu^\gamma.
\end{equation}
For a single generation of stars formed at $t=0$, the number of 
SNe Ia explosions up to time $t$ is then given by
\begin{equation}
R_{\rm Ia}(\le t) 
= B \int_{m_{\rm B, low}}^{m_{\rm B, up}} \phi(m_{\rm B}) 
\left\{\int_{\mu_{\rm min}(t)}^{1/2} f(\mu) \dm \mu\right\}
\frac{\dm m_{\rm B}}{m_{\rm B}}, 
\end{equation}
where the lower limit
\begin{equation}
\mu_{\rm min}(t) = {\rm max} \left\{ \frac{M(t)}{m_{\rm B}}, 
\frac{m_{\rm B} - m_{\rm B, up}/2}{m_{\rm B}}\right\}
\end{equation}
is set by the conditions that the secondary has evolved off the main 
sequence and that $m_1 = m_{\rm B} - m_2 \leq m_{\rm 1, max}$. 
Following Portinari et al. (1998), 
we adopt $m_{\rm B, low} = 3 M_{\odot}$, $m_{\rm B, up} = 12 M_{\odot}$, 
$\gamma = 2$, and $B=0.07$. 
Finally, the yield of metals from SNe Ia is computed as 
\begin{equation}
p_{{\rm Ia}, i}(\le t) 
= M_i^{\rm Ia} R_{\rm Ia}(\le t), 
\end{equation}
where $M_i^{\rm Ia}$ is taken from the W7 model of \citet{nom97}. 
In our code, we follow the evolution of oxygen, iron, magnesium, and 
silicon. 
Since we will vary the IMF, the self-consistent treatment of metals 
is very important, and allows us to make predictions,
for example, for the abundance and abundance ratios 
of the $\alpha$-elements to the iron-peak elements. These aspects
of the simulation will be considered in a later paper.

As described above, each stellar particle returns mass, energy, and
metals to the ambient gas according to its age and metallicity.  We
smoothly distribute these quantities amongst the neighbouring gas
particles using the SPH smoothing. There are various possible ways to
define the smoothing length of a stellar particle. As the masses of
SPH particles are variable, we choose to set the smoothing length of a
stellar particle so as to have a constant mass (rather than number)
within the smoothing volume.  This method ensures numerical
convergence if one chooses the constant mass as $M_{\rm ngb} =
\alpha m_{\rm orig}$, where $M_{\rm ngb}$ is the gas mass within the
smoothing volume and $\alpha$ is a parameter whose value should not be
changed by the resolution. In general, smaller $\alpha$ results in
stronger feedback effects as expected.  As a rational choice we use the same
mass resolution as in the hydrodynamic calculation, that is $\alpha = 40$.
The mass ejected from evolved stars and SNe Ia is simply added to the
mass of the hot phases of the neighbouring gas particles.  For an SPH
particle that receives an ejected mass $\Delta M_{\rm FB}$ during the
timestep, we assign a new effective specific energy $u_{\rm eff}'$ and 
a new density $\rho'$ according to
\[u_{\rm eff}' 
	= u_{\rm eff} \frac{M_{\rm SPH}}{M_{\rm SPH} + \Delta M_{\rm
	FB}} 
\] 
and 
\[\rho' = \rho \ \frac{M_{\rm SPH} + \Delta M_{\rm FB}}{M_{\rm SPH}}.\] 

Using equation (\ref{EV}), the mass evaporated by SNe is calculated 
according to the energy, $\Delta Q_{\rm FB}$, 
received during the timestep $\Delta t$ :
\begin{equation}
\Delta M_{\rm EV} = A \frac{\Delta Q_{\rm FB}}{u_{\rm SN}},  
\end{equation}
where we compute $u_{\rm SN}$ considering only contributions from stars 
heavier than $8 M_{\odot}$ using the IMF for the quiescent star formation 
mode (see \S \ref{quiescent}).  

The new mass of the hot phase, $M_{\rm h}'$, is given by solving the thermal 
energy equation implicitly, using eqn.~(11):
\begin{equation}
M_{\rm h}' =  M_{\rm h} + \Delta M_{\rm EV}
-\frac{\Lambda_{\rm net}(\rho_{\rm h}', u_{\rm h}, Z)}
	{u_{\rm h}-u_{\rm c}}\frac{M_{\rm h}'}{\rho_{\rm h}'} \Delta t, 
\end{equation}
where the new density of the hot phase, 
$\rho_{\rm h}' = \rho_{\rm h}(M_{\rm h}')$,  
is given by equations (\ref{dens}), (\ref{mass}), and (\ref{pequil}) 
assuming $u_{\rm h}$ and the particle volume remain constant during 
cloud growth. 

The new specific energy of the hot phase, $u_{\rm h}'$, is also calculated 
implicitly, 
\begin{equation}
u_{\rm h}' = u_{\rm h} + 
	\frac{\Delta Q_{\rm FB} - (u_{\rm c} - u_{\rm h}')\Delta M_{\rm EV}}
	{M_{\rm h}'}. 
\end{equation}

We impose an additional timestep criterion, $\Delta t_{\rm FB}$, 
for stellar particles so as to capture the non-instantaneous feature 
properly: 
\begin{equation}
\Delta t_{\rm FB} = {\rm max}(c_{\rm FB} t_{\rm age}, 0.5 {\rm Myr}), 
\end{equation}
where $c_{\rm FB}$ is a parameter and $t_{\rm FB}$ is the age of a 
stellar population. We have confirmed that the simulation results 
converge for $c_{\rm FB} < 0.1$, and thus employ $c_{\rm FB} = 0.05$ 
throughout.

\subsubsection{Quiescent star formation mode and parameter setting} 
\label{quiescent}

The model described above has three parameters: the threshold
density, $\rho_{\rm th}$, the evaporation efficiency parameter $A_0$,
and the characteristic star formation timescale $t_*^0$.  Before we
can specify these parameters, we have to fix the IMF and the amount of
energy released by a SN.  For this quiescent mode of star formation,
we use the Salpeter IMF \citep[$x = 1.35$ in
eq. (\ref{IMF});][]{sal55} with $m_{\rm u} = 100 M_{\odot}$ and
$m_{\rm l} = 0.1 M_{\odot}$ and adopt $2 \times 10^{51}$ erg for the
energy released per SN, which yields a SN temperature $T_{\rm SN} = 2
\mu u_{\rm SN}/(3k) \simeq 3.3 \times 10^8$ K, where $\mu$ denotes the
mean molecular weight.  The reason why we use this rather than the
canonical value of $10^{51}$ erg per SN is simply that we need this
amount of energy in order to reproduce the Kennicutt (1998) law for
gas with solar metallicity.  For primordial gas, the value $10^{51}$
erg/SN is sufficient.  Note that an IMF with a single slope is too
simple a description \citep[e.g.][]{ken83, kro98} and changing the
slope of the low mass portion, say for $m < 1 M_{\odot}$, can easily
change the number of SNe II by a factor of $\sim 2$.  For simplicity,
we restrict ourselves to using a single-slope IMF, but we reserve the
freedom to increase the energy per supernovae just discussed.

To fix the model parameters, we considered self-regulated star
formation in a self-gravitating gas sheet, adopting the instantaneous
recycling approximation. This procedure is identical to that followed
by \citet{sh03}, so we will give no further details here.  The values
of the parameters we adopt are given in Table~\ref{param}.

\begin{table}
\caption{
The values of the model parameters, adjusted  to reproduce 
the Kennicutt (1998) law represented by eqn~(25) in \citet{sh03}. } 
\label{param}
\begin{center}
\begin{tabular}{@{}ccc}
\hline
  $\rho_{\rm th}$ &
  $t_*^0$ &
  $A_0$ \\
\hline 
  $ 2 \times 10^{-25}$ g cm$^{-3}$ & 
  2.8 Gyr & 
  1300\\
\hline
\end{tabular}
\end{center}
\end{table}
\begin{figure} 
\includegraphics[width=7.5cm]{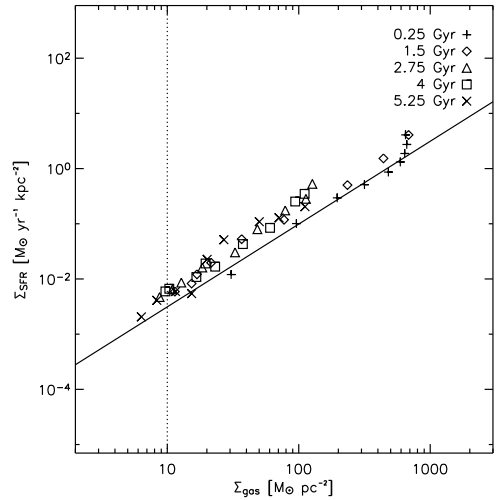} 
\caption{Star formation rate per unit area versus gas surface density
in an idealised simulation of disc formation in a virialised halo.
The surface star formation rate densities are computed using the
surface density of stars younger than $3 \times 10^{7}$ yr in
cylindrical bins. The outermost bin corresponds to the edge of the star
forming region.  The symbols represent star formation rates at
$t=0.25, \ 1.5, \ 2.75, \ 4, \ 5.25$ Gyr, indicated by crosses, diamonds,
triangles, squares, and crosses respectively.  The solid and dotted
lines indicate the target relation and the cut-off surface density
respectively.}  
\label{kennicutt} 
\end{figure} 
In Fig.~\ref{kennicutt}, we show the relation between the star
formation surface density and the gas surface density in an idealised
simulation of disc formation in a virialised halo
\citep[see][]{oka03}.  Even though the values of the parameters were
set assuming the instantaneous recycling approximation and solar
metallicity, this simulation, which did not make this approximation
and followed chemical evolution, shows reasonable agreement with the
target relation.  The surface gas density at the edge of star-forming
region is also consistent with the threshold surface density.

\subsubsection{The burst mode and top-heavy IMF}
 
In the self-regulated star formation that occurs in the quiescent star
formation mode, cooling and feedback are balanced.  We implement a
second mode of star formation, which is able to blow gas out of
galaxies.  For this to happen, the injection of feedback energy has to be
more rapid. Hence, we require that star formation should occur in short
{\it bursts} in physical conditions that permit heating by feedback to
exceed local cooling.  After various tests, we found that sufficient
heating cannot be obtained simply by adopting a shorter $t_0^*$ (or
equivalently, a larger $c_*$).  Motivated by the considerations
discussed in \S~1, we therefore decided to assume a top-heavy IMF
in the burst mode as well as a short star formation timescale.  We use
$t_0^* \simeq 0.15$ Gyr which corresponds to $c_* = 0.5$ and an IMF
with slope $x = 0.34$, which
maximises the number of SNe II.  The other parameters including the
upper and lower mass limits for the IMF are the same as for quiescent
star formation.

In semi-analytic galaxy formation models, it is usually assumed that a
starburst is triggered by a major galaxy merger \citep[e.g.][]{wf91,
kau93, col00}, although some studies include small starbursts induced
by minor mergers as well \citep{sp99, on03}.  In our simulations, we
would like to model the burst on the basis of local physical 
quantities rather than on global information such as the merger mass
ratio,  which is not easily available.  Exactly how to do this, however,
is not evident. In this paper, we examine two possibilities for
triggering the burst mode of star formation.

Firstly, we consider  the gas density. \citet{mh94} showed that
cold gas is driven to the centre of the remnant when galaxies merge.
The burst is therefore preceded by an increase in the central density
and it seems plausible that imposing a threshold density, $\rho_{\rm
burst}$, above which the burst mode is switched on, can capture the 
conditions under which nuclear starbursts occur in merging galaxies.  In
addition, \citet{sh03} noted that self-regulation in the multi-phase
model breaks down at sufficiently high density; our ``density-induced"
burst may be considered as a crude modelling of this process.

The second plausible trigger for bursts is the presence of shocks.
Galaxy mergers induce strong, galactic-scale shocks. Indeed,
\citet{bar04} suggested that shock-induced star formation is the
dominant mode in interacting galaxies.  We model a shock-induced burst
as follows. When the rate of change of the entropy variable $K(s)
\equiv (\gamma-1) u \rho^{1-\gamma}$ due to the artificial viscosity
exceeds a threshold value, $\dot{K}_{\rm burst}$, the burst mode is
switched on.  We use $\gamma = 5/3$ throughout.  Note that other
variables such as $\dot{u}$ or $\rho/(\dot{\rho}\ t_{\rm dyn})$ can also
be used to identify shocked particles, and we have confirmed that
these alternative choices produce the same results for appropriate
values of the thresholds.  In contrast to the density-induced burst, the
shock-induced burst is expected to result in a large scale (galactic
scale) starburst in interacting galaxies. The shock trigger is more
sensitive to mergers than the density trigger because in the latter
case, the burst cannot start until sufficient gas has been funneled
into the galactic centre by the merger \citep{bar04}.

We use $\rho_{\rm burst} = 10^{-23}$ g cm$^{-3}$ and $\dot{K}_{\rm
burst} = 9 \times 10^3$ in our code units ($h^{-1}$ Mpc, km s$^{-1}$,
and $10^{10} h^{-1} M_{\odot}$ for length, velocity, and mass,
respectively) throughout this paper.  The dependence of our results on
these parameters will be discussed later.  It should be noted that,
for the density-induced burst, our results do not depend on resolution
as long as this is good enough to resolve the threshold density
$\rho_{\rm burst}$.  On the other hand, the shock-induced burst will
have a strong dependence on resolution because the width of the
shocked layer is simply proportional to the smoothing length. Thus, we
would need to adjust $\dot{K}_{\rm burst}$ or $t_0^*$ for the burst
mode if we were to change the numerical resolution.

\section[]{SIMULATION SETUP}

In order to study disc formation, we have selected from a pre-existing
cosmological $N$-body simulation a halo with a quiet merger history.
The semi-analytic model of \citet{col00} applied to the merger tree of
this halo, predicts a galaxy that is disc-dominated at the present
day.  This halo is, in fact, the same as the one used in
\citet{oka03}. In that paper, we showed that, if cooling is not
allowed to occur until $z = 1$, then a reasonable disc galaxy forms by
$z = 0$.

\subsection{Initial condition}

We assume a low-density, flat CDM universe ($\Lambda$CDM) as the
background cosmology. We adopt the following choices for the
cosmological parameters: mean matter density, $\Omega_0 = 0.3$, Hubble
parameter, $h \equiv H_0/100 {\rm km s}^{-1} {\rm Mpc}^{-1} = 0.7$,
cosmological constant term, $\Omega_\Lambda \equiv \Lambda_0/(3 H_0^2)
= 0.7$, amplitude of mass fluctuations, $\sigma_8 = 0.9$, and mean
baryon density, $\Omega_{\rm b}=0.04$

To generate our initial conditions, we use the resimulation technique
introduced by \citet{fews96}. The halo of interest was grown in a dark
matter simulation of a 35.325 $h^{-1}$ Mpc periodic cube.  Its mass is
about $M_{\rm vir} = 1.2 \times 10^{12} h^{-1} M_{\odot}$ within the
sphere which has the virial overdensity, $\delta_{\rm vir} = 337$, at
$z = 0$.  The circular velocity, spin parameter, and collapse redshift
of this halo are $v_{\rm c}(r_{\rm vir}) = 155$ km s$^{-1}$, $\lambda
\equiv J |E|^{1/2}/(G M^{5/2}) = 0.038$ and $z_{\rm c} \simeq 1.5$,
respectively, where $z_{\rm c}$ is defined as the redshift at which
the main progenitor had half the final halo mass.  To generate the new
initial conditions, the initial density field of the parent simulation
is recreated and appropriate additional short wavelength perturbations
are added in the region out of which the halo forms.  In this region
we also place SPH particles in the ratio of 1:1 with dark matter
particles. The region external to this was populated with high-mass
dark matter particles, the function of which is to reproduce the
appropriate tidal fields. The initial redshift of the simulation is
50. The masses of the SPH and high-resolution dark matter particles
are $\sim 2.6 \times 10^6 h^{-1} M_{\odot}$ and $\sim 1.7 \times 10^7
h^{-1} M_{\odot}$, respectively.

The gravitational softening lengths are kept fixed in comoving
coordinates for $z > 3$; thereafter they are frozen in physical units
at a value (of the equivalent Plummer softening) of $\epsilon=0.5$ kpc
and $1$ kpc for the SPH and high-resolution dark matter particles
respectively.  The gravitational force obeys the exact $r^{-2}$ law at
$r > 2.8 \epsilon$.

\subsection{Models}

To investigate the effects of the burst mode of star formation on the
evolution of the galaxy, we consider three models. In the first, we do
not include a burst mode. We refer to this reference case as the {\it
no-burst} model.  The second (the {\it density burst} model) has the
density-induced burst mode and the third (the {\it shock-burst}
model), the shock-induced burst mode.  To study the effects of the
burst mode in detail, we also consider, in \S\ref{dependence}, models
with different threshold values for triggering the burst, as well as
models with a combination of density-induced and shock-induced bursts.

\section{RESULTS}

First, we present a description of the final galaxy that forms in each
simulation with a different feedback prescription, and then undertake
a detailed study of the evolution in each case. In the final
subsection, the burst parameters are varied in order to test the
sensitivity of the simulation to these assumptions.

\subsection{Galaxy properties at $z=0$} 
\begin{figure*} 
\includegraphics[height=22cm]{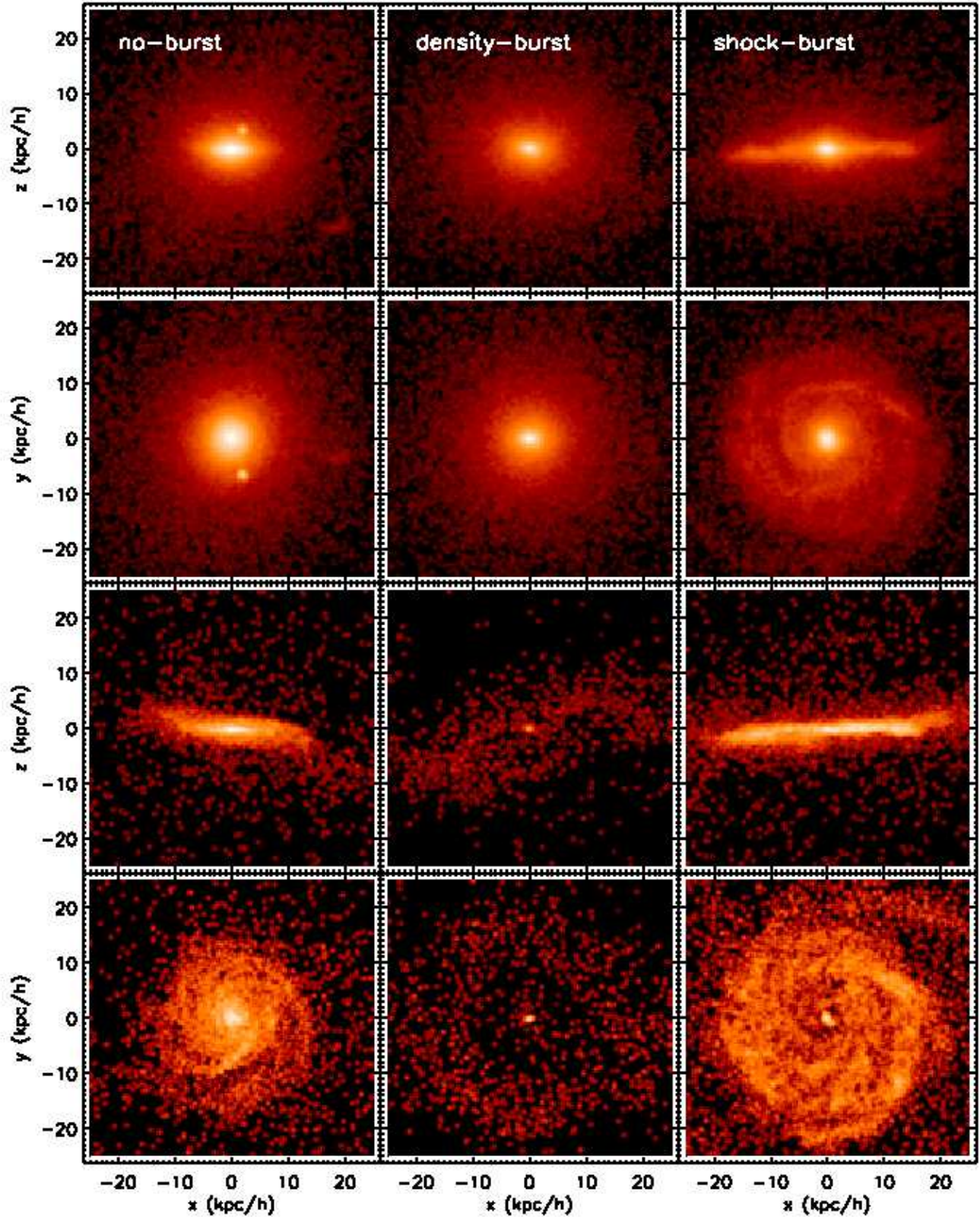} 
\caption{ Galaxies at $z=0$. The
no-burst, density-burst, and shock-burst models are shown from left to
right.  The edge-on and face-on views of stars, and edge-on and
face-on views of gas are given in rows from top to bottom.  We use the
nett angular momentum of stars within 10 $h^{-1}$ kpc spheres to
define the viewing angles. The brightness indicates the projected mass
density, and the same scaling is used for each model.  All of these
galaxies are obtained from the same initial conditions.}
\label{z0}
\end{figure*} 
Fig. \ref{z0} shows the stellar and gas distributions within 50
$h^{-1}$ kpc boxes centred on the galaxies at $z=0$. The edge-on and
face-on projections are selected to be perpendicular and parallel to
the angular momentum vector of the stellar component within the
central 10 $h^{-1}$ kpc sphere.  The no-burst galaxy has a tiny
stellar disc and a large population of halo stars. Visually, this
galaxy is similar to the one that \citet{aba03a} obtained in their
simulation. It should be noted that our effective equation of state
for the ISM is as stiff as that used by \citet{rob04}, and hence the
stabilisation of gas against the Toomre instability is insufficient to
allow a large stellar disc to form in this particular halo.  The
simulation with the density-burst model produces an ellipsoidal
stellar object that is slowly rotating. Due to strong feedback in the
galactic centre arising from the density-induced starbursts, most of
the gas is expelled from the galaxy and only a diffuse gaseous ring
remains.  In contrast, in the shock-burst model the number of halo
stars is significantly lower and an extended disc forms in both the
stellar and gaseous components.  At the galactic centre, dense gas
cores always exist regardless of the model. We suspect that this is an
artificial effect caused by our limited resolution.

\begin{figure} \includegraphics[width=7.5cm]{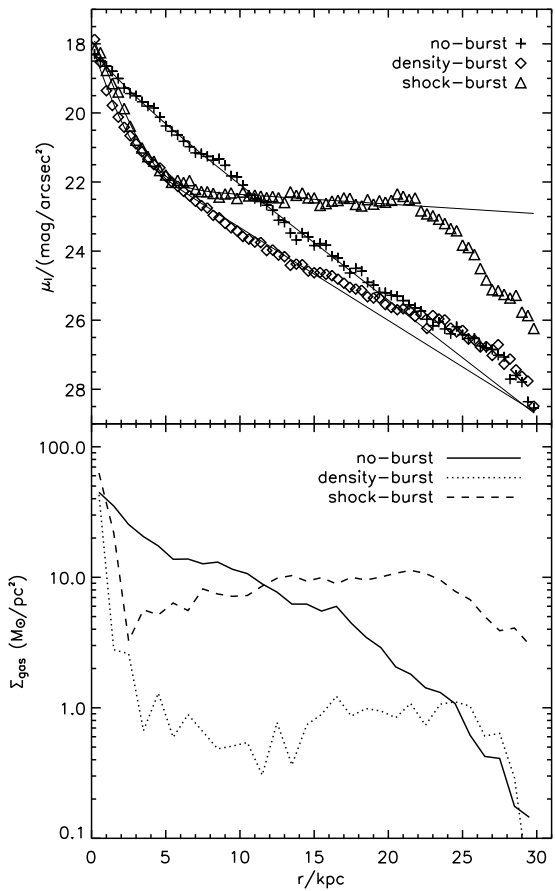} 
\caption{
Upper panel: Surface brightness profiles in the $I$-band for the no-burst 
(crosses), density-burst (diamonds), and shock-burst (triangles) models. The
solid lines are double-exponential fits for $r < 20$ kpc and 
the fitting parameters are given in Table~\ref{fitparam}. 
Lower panel: Cold gas surface density profiles. The solid, dotted, and dashed lines 
represent the no-burst, density-burst, and shock-burst models, respectively. 
Note that we have removed the $h$ dependence here.}  
\label{surface} 
\end{figure} 
The $I$-band surface brightness profile of each galaxy is shown in
the upper panel of Fig.~\ref{surface}. To compute luminosities, 
we used the population synthesis code P\'{E}GASE2
\citep{pegase}. The luminosity of each particle is calculated for the
IMF, metallicity and age appropriate to that particle in the
simulation. We fit each surface brightness profile up to 20 kpc with a
double-exponential because the standard $r^{1/4}$ + exponential
profile cannot fit the shock-burst galaxy. It should be noted,
however, that the $r^{1/4}$ + exponential profile is a better
description for the density-burst model, and therefore the double-exponential 
fit underestimates its bulge. Values of the fitting
parameters are shown in Table
\ref{fitparam} which also gives the total $B$- and $I$-band
luminosities of each galaxy. 

The surface brightness profile of the no-burst galaxy in
Fig.~\ref{surface} is well fit by a single exponential (90 \% of the 
$I$-band light comes from the outer exponential disc). 
Thus, there is no evidence in this profile for a bulge component 
even though Fig.~\ref{z0} clearly shows an extended spheroidal component 
around this galaxy. 
The double-exponential is a good fit to the shock-burst
galaxy, but the outer exponential is rather flat, indicating that the
surface brightness of the disc is nearly constant as a function of
radius. The resulting scale length, 39 kpc, is therefore very large. As
we discuss in \S5, it is possible that the disc acquires excessive
angular momentum because the galactic winds are not collimated and
exert a pressure on the hot gas reservoir which becomes distended and
more susceptible to tidal torques at early times. In spite of its large
size, the shock-burst galaxy falls on the $I$-band Tully-Fisher
relation. For its rotation velocity of $v_{\rm rot}
\simeq 200$ km s$^{-1}$, its absolute magnitude, $M_{I} = -22.7$, is
only slightly brighter than the mean relation and well within the
scatter \citep{giov}.

The surface brightness profile in each galaxy reflects the cold gas surface  
density profile (lower panel). In the no-burst galaxy, the cold gas 
disc has an approximately exponential profile that is slightly more extended 
than the distribution of the stellar light. The density-burst galaxy has lost 
its gas content due to strong winds as we will show in \S4.2. As a result, 
the gas surface density is an order of magnitude lower than the Kennicutt 
threshold ($\Sigma_{gas} \simeq 10 M_{\odot}$ pc$^{-2}$). 
In the shock-burst model, a plateau is also seen in the gas surface  
density profile, and the surface density is close to the Kennicutt threshold. 
The three different models all have similar central surface gas densities. 
This implies that the central gas cores are a numerical artifact 
and higher resolution or kinetic feedback may be required to get rid of these 
cores.

\begin{table} 
\caption{Values of the parameters used in Fig.~\ref{surface} 
to fit the $I$-band surface brightness profiles.  The central surface
brightnesses $\Sigma^0$ and the scale lengths $R^d$ are given for
the inner and outer profiles.  Subscripts $i$ and $o$ indicate the
inner and outer profiles respectively. The last 2 columns give the
total $B$- and $I$-band luminosities for each galaxy.}  
\label{fitparam}
\begin{center} 
\begin{tabular}{@{}lcccccc} 
\hline
  & $\Sigma^0_{\rm i}$
  & $\Sigma^0_{\rm o}$
  & $R^{\rm d}_{\rm i}$/kpc 
  & $R^{\rm d}_{\rm o}$/kpc 
  & ${\rm M_B}$
  & ${\rm M_I}$\\
\hline 
no-burst & 19.1 & 18.8  & 1.5  & 3.3    & -21.1 & -22.6\\
density-burst & 19.1 & 20.6 & 0.8 & 4.0 & -19.7 & -21.4\\
shock-burst & 17.7 & 22.1 & 1.0 & 39 & -21.3 & -22.7\\
\hline \end{tabular}
\end{center}
\end{table} %

\begin{figure}
\includegraphics[width=7.5cm]{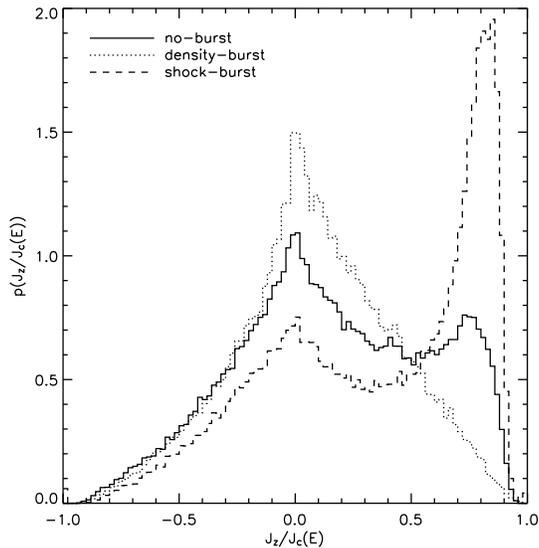}
\caption{
Mass-weighted probability distributions of the orbital circularity,  
$J_z/J_{\rm c}(E)$, within 25 $h^{-1}$ kpc from the galactic centre.  
The solid, dotted, and dashed lines indicate the no-burst, density-burst, 
and shock-burst models respectively. 
}
\label{decomposition}
\end{figure}

The fits to the surface brightness profiles of Fig.~\ref{surface}
provide only partial information about the morphology of the galaxy. A
more informative way to characterise the relative importance of bulge
and disc components is to carry out the dynamical decomposition
proposed by \citet{aba03b}.  For this, we first compute the angular
momentum, $J_z$, of each star particle parallel to the nett angular
momentum of stars within $10 h^{-1}$ kpc, and the angular momentum of
the co-rotating circular orbit, $J_{\rm c}(E)$. The ratio $J_z/J_{\rm
c}$ defines an orbital circularity. In Fig.\ref{decomposition}, we
show the probability distribution of this orbital circularity for
stars within 25 $h^{-1}$ kpc of the galactic centre. A disc component
should have $J_z/J_{\rm c}(E) \simeq 1$ and such a component is
clearly visible in the shock-burst model, confirming the visual
impression gleaned from Fig.~\ref{z0}. By assuming a non-rotating
spheroid, i.e. that stars in the spheroid are symmetrically
distributed around zero, all stars having $J_z/J_{\rm c}(E) \leq 0$
are identified as a half of the spheroid. Their counterparts with
$J_z/J_{\rm c}(E) > 0$ are defined statistically. All remaining stars
are identified as the disc component. We do not try to decompose the
disc into thick and thin discs but defer a detailed study to a
forthcoming paper.

In Table.~\ref{bt}, we show the disc-to-total ratios for mass, $U$,
$B$, $V$, $I$, and $K$-bands.  The redder the band, the more the
spheroid dominates.  These ratios confirm that the shock-burst model
has the most significant disc, while the density-burst model is the
most spheroid-dominated. The $B$-band disc-to-total luminosity ratio
of the shock-burst galaxy is $D/T = 0.84$, which is sufficiently large
to be identified conventionally as a disc galaxy, while the relation 
between the $D/T$s and the Hubble T-types has a large scatter 
\citep{bau96, gra01}.  
Note that the dynamical decomposition reveals a significant spheroidal 
component in the no-burst galaxy even though the photometric decomposition
based on the surface brightness profile failed to detect it. 

\begin{table}
\caption{Disc-to-total mass and luminosity ratios for the simulated 
galaxies at $z=0$.}
\label{bt}
\begin{center}
\begin{tabular}{@{}lcccccc}
\hline
  & mass &
  $U$ &
  $B$ &
  $V$ & 
  $I$ & 
  $K$ \\
\hline 
no-burst & 0.26 & 0.72 & 0.63 & 0.54 & 0.45 & 0.44 \\
density-burst & 0.21 & 0.22 & 0.22 & 0.22 & 0.21 & 0.20 \\
shock-burst & 0.48 & 0.86 & 0.84 & 0.80 & 0.72 &0.66 \\
\hline
\end{tabular}
\end{center}
\end{table}

\subsection{Mass accretion and star-formation histories}\label{evolution}

\begin{figure}
\includegraphics[width=7.5cm]{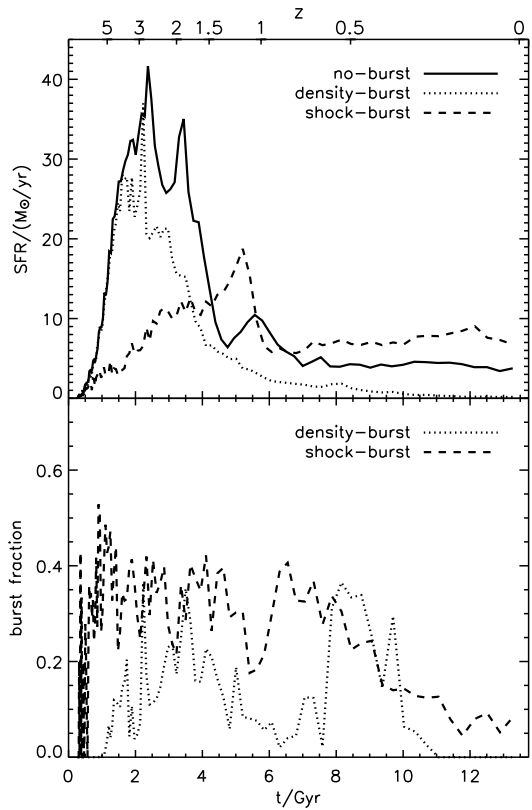}
\caption{
Upper panel: the formation history (star formation rate as function of
time) of the stars that lie within 25 $h^{-1}$ kpc from the galactic
centre at $z = 0$. Lower panel: the fraction of stars formed in the
burst mode.  The solid, dotted, and dashed lines correspond to the
non-burst, density-burst, and shock-burst models respectively. }
\label{sfr}
\end{figure}

The star formation histories (SFHs) of the simulated galaxies are
plotted in Fig.~\ref{sfr}. In the no-burst simulation, significant
star formation activity takes place at high-redshift, and this early
star formation is responsible for the massive bulge (and halo)
population that forms in this model.  This galaxy has a nearly constant
star formation rate after $z=1$, and it builds up the tiny
disc present at the final time.

The SFH of the density-burst model is similar to that of the no-burst
model up to the point when the gas density reaches the threshold for a
burst.  At this point, the burst is triggered, the associated feedback
suppresses further star formation, and the star formation rate falls
well below that in the no-burst model. The huge amount of feedback
energy released into a small region near the centre of the galaxy
blows out most of the hot gas and this almost completely quenches
further star formation. As a result, the density-induced burst
simulation yields an elliptical-like galaxy.  If the threshold is
lowered, then the burst is turned on earlier and even fewer stars
form. Thus, it is difficult to produce a galaxy that is more
disc-dominated than in the no-burst case simply by allowing
density-induced bursts.

By contrast, shock-induced bursts occur even at very high redshift, in
small haloes where bursts are triggered both by mergers and violent
collapse. The strong feedback from the stars formed in these bursts
suppresses cooling and star formation during this early period when
the galaxy is undergoing a number of merger events. As a result, the
peak in the SFH shifts to $t \sim 5.3$ Gyr ($z \sim 1.1$).  An
extended reservoir of hot gas is created by the feedback energy
associated with the burst and much of it remains attached to the
halo. After $z=1$, when the bulk of the dynamical activity is over,
this gas cools to form an extended , young stellar disc. The key features
of the shock-induced burst may be seen in the lower panel of
Fig.~\ref{sfr}. The fraction of stars born in the burst mode is large
($\sim 40$ per cent) at high-redshifts ($t < 6$ Gyr) and it gradually
decreases to $\sim 5$ per cent towards the present.  Thus, the
shock-burst model experiences strong feedback at early times, followed
more recently by a sustained period of steady, self-regulated star
formation.

The top-heavy IMF present in bursts delivers feedback energy in a
particularly efficient way. It is worth noting that even at its peak,
the star formation rate in the shock-burst model barely reaches $\sim
20 {\rm M_\odot/yr}$. However, the star formation rate in the
starburst would be significantly overestimated if, as is the norm in
observational studies, the rates are inferred by assuming a standard
IMF.

\begin{figure}
\includegraphics[width=7.5cm]{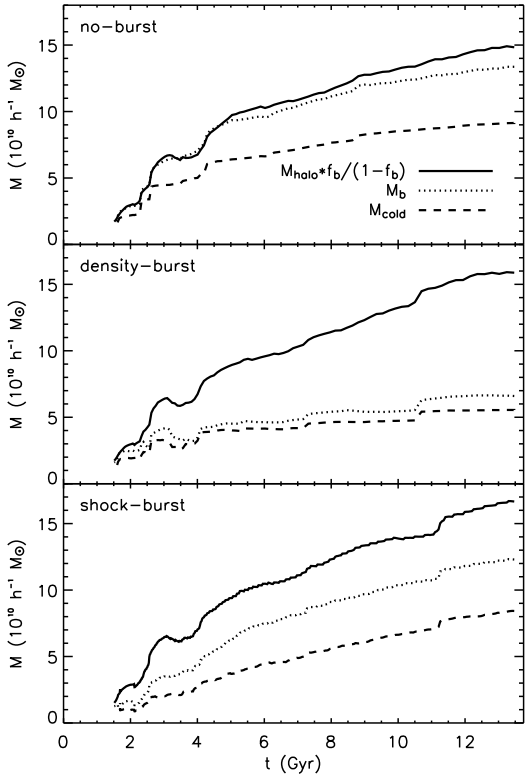}
\caption{
Evolution of the mass in the main progenitor haloes. The solid, dotted,
and dashed lines represent the normalised halo mass, baryon mass, and
the cold baryon (cold gas and stars) mass in the halo,
respectively. We show the no-burst, density-burst, and shock-burst
models from top to bottom. }
\label{massevolution}
\end{figure}
To investigate the build-up of the visible components of the galaxy
and the importance of galactic winds, we now consider the mass
evolution of the main progenitor of the halo as a function of 
time. We define the mass of the main progenitor as $M_{\rm tot} =
(4\pi/3) r_{\rm vir}^3 \rho_{\rm vir}(z)$ where $r_{\rm vir}$ is given
by the spherical collapse model in our chosen cosmology
\cite{ecf96}. Since the accretion and cooling of baryonic matter
affects the structure of the dark matter halo, the halo mass defined
in this way differs slightly between models: the more concentrated the
dark matter is, the smaller the halo mass.  In
Fig.~\ref{massevolution}, we plot the normalised main progenitor halo
mass, $f_{\rm b}M_{\rm halo}/(1-f_{\rm b})$, where $f_{\rm b} \equiv
\Omega_{\rm b}/\Omega_0$ is the baryon fraction in the universe. Also
plotted are the mass of baryonic matter, $M_{\rm b}$, and the mass of
cold baryons (stars and cold gas), $M_{\rm cold}$, in the main
progenitor halo. If the halo had the mean cosmic baryon fraction,
$f_{\rm b}$, then the normalised halo mass would be equal to the
baryon mass.

The baryon fraction in the no-burst galaxy is always very close to the
cosmic value. Very little gas is lost from the halo in this simulation
because all the feedback energy goes mostly to support the ISM against
its self-gravity. In the density-burst simulation, the baryon fraction
in the halo is quite low and there is almost no hot gas left,
signaling the existence of strong galactic winds. The stepwise jumps 
in the cold baryon mass indicates that this mass grows primarily
through mergers with other haloes.  The shock-burst galaxy also has
winds, but these are not as strong as in the density-burst model. By
the final time, the halo of the shock-burst galaxy has lost 24 per
cent of its baryons. While the cold baryon content at $z=0$ is similar
to that in the no-burst simulation, the rate of growth of the cold
baryons is steeper at early times in the no-burst model and at late
times in the shock-burst model. This is the origin of the difference
in the morphology of the final galaxy in these two cases.

\begin{figure*}
\includegraphics[height=22cm]{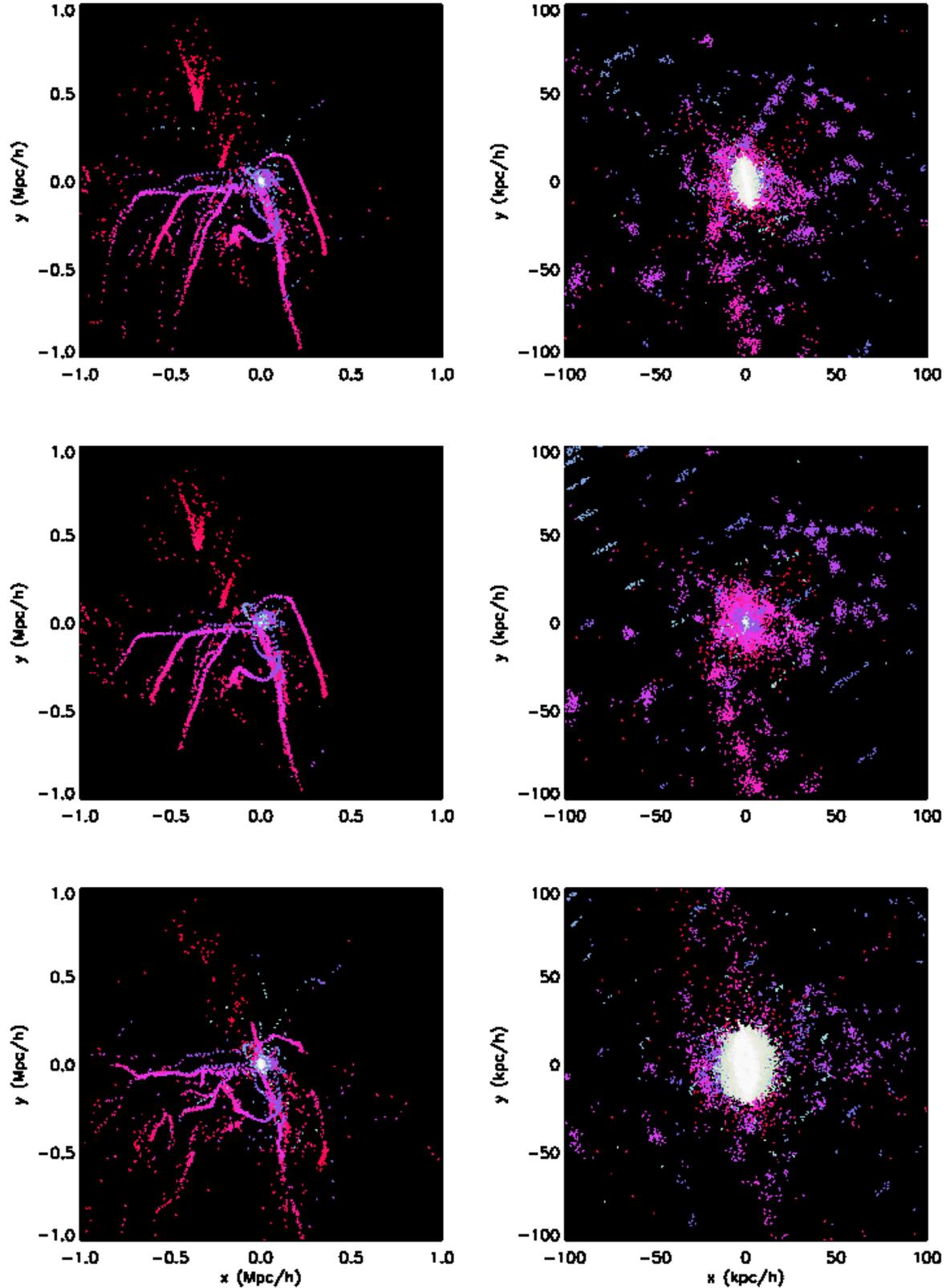}
\caption{
Birthplaces of stars that lie within $25 h^{-1}$ kpc of the $z=0$
galactic centre. These birthplaces are plotted relative to the
position of the centre of the main progenitor at the epoch when each
star forms. The colours indicate the formation time of the stars, with
red being the oldest and white the youngest. Panels show the no-burst
(top row), density-burst (middle), and shock-burst (bottom) models.
The left and right panels show the $x$--$y$ projection of the
birthplaces in $2 h^{-1}$ Mpc and $200 h^{-1}$ kpc cubes respectively.
The coordinate system used here is that of the original simulation,
i.e. it has not been rotated to match the angular momentum of the
final galaxy.  }
\label{birthplace}
\end{figure*}
The behaviour of both the SFH and the mass evolution in the
shock-burst model indicates that the success of this model stems from
the suppression of star formation in small haloes. In order to identify
the location of the main star formation sites, we plot, in
Fig.~\ref{birthplace}, the birthplaces, relative to the position of
the main progenitor at the time, of the stars that end up within $25
h^{-1}$ kpc of the galactic centre at $z=0$.  The continuous
trajectories of the birthplaces in the no-burst model (top left panel)
indicate that small haloes undergo continuous star formation until they
fall into the main progenitor. A similar behaviour is seen in the
density-burst model. Since density-induced bursts do not occur in
small haloes, where the gas density cannot reach the threshold value,
star formation in such small systems is the same as in the no-burst
model. The main difference between the two models is apparent in the
zoomed panel on the right-hand side of the figure: strong feedback
from density-induced bursts almost stops star formation in the main
progenitor after $z \sim 1$. The main sites of star formation in this
model are small infalling haloes. This is how the elliptical galaxy
forms in the density-burst model. Unlike density-induced bursts,
shock-induced bursts can occur in small haloes whose shallow potential
wells make feedback particularly effective. The star formation
trajectories are therefore no longer continuous. Instead, intermittent
star formation in infalling haloes is clearly evident in the lower
left-hand panel. As anticipated, the suppression of star formation in
small galactic building blocks is the key to making a large disc.

More quantitative information on the location of the star formation
sites is provided in Table~\ref{ratios} which lists the fraction of
the final stellar mass within $25 h^{-1}$ kpc from the galactic centre
that has formed within a comoving sphere of radius $25 h^{-1}$ kpc
around the main progenitor.  The mass of each star used in this
calculation is the mass at $z=0$, not the initial mass.  We find that
78 per cent of the stellar mass formed in the main progenitor in the
shock-burst model. In contrast, this fraction drops to 25 per cent in
the case of the density-burst model.
\begin{table}
\caption{The stellar mass formed within a comoving 
$25 h^{-1}$ kpc sphere around the centre of the main progenitor
divided by the total stellar mass within $25 h^{-1}$ kpc from the
galactic centre at $z=0$. }
\label{ratios}
\begin{center}
\begin{tabular}{@{}ccc}
\hline
  no-burst & density-burst & shock-burst \\
\hline
  0.60 & 0.25& 0.78 \\
\hline
\end{tabular}
\end{center}
\end{table}
\begin{figure*}
\includegraphics[height=22cm]{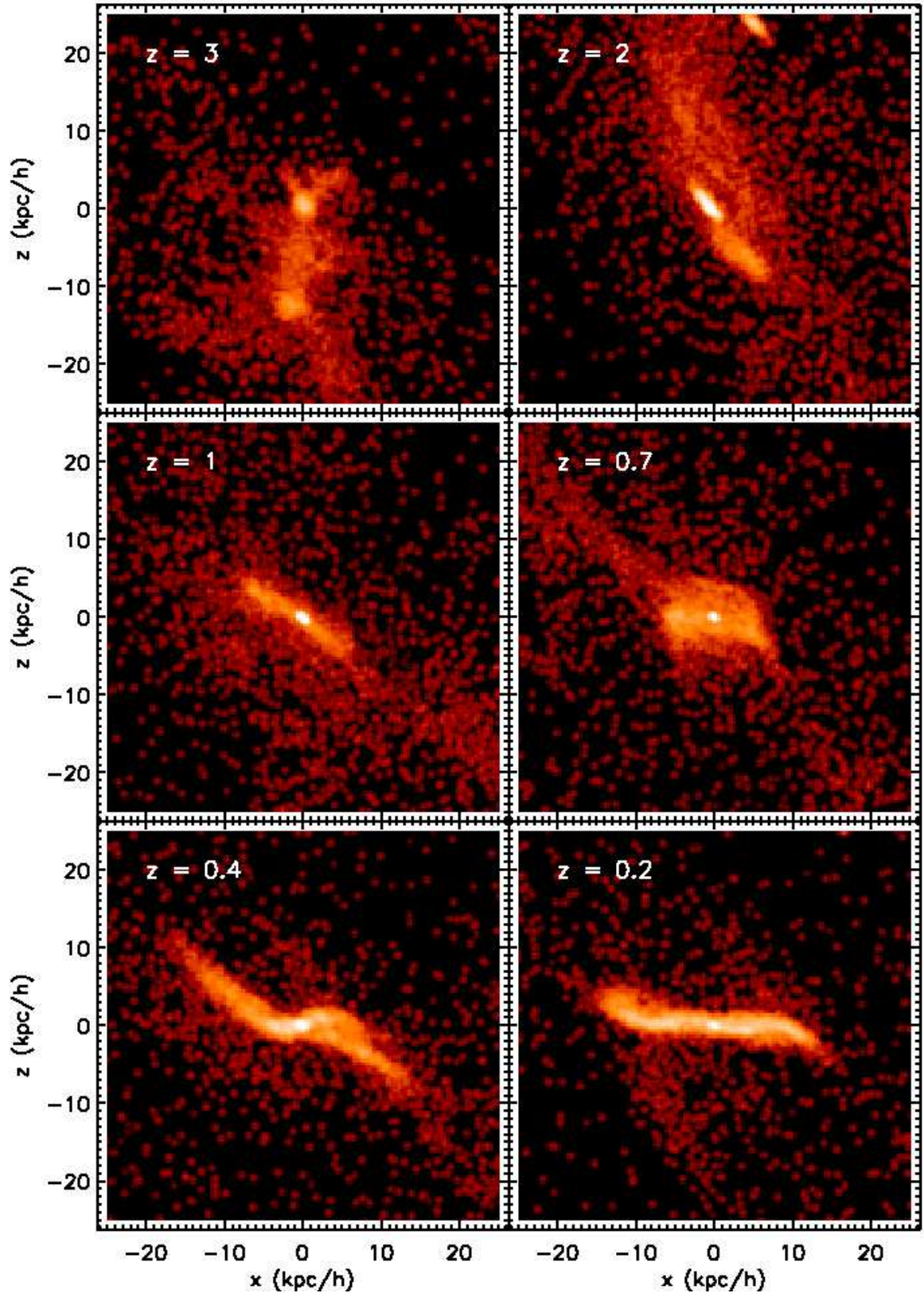}
\caption{
Gas distribution in physical $50 h^{-1}$ kpc boxes centred on the main
progenitors in the shock-burst model. A fixed viewing angle is used,
that for which the disc is edge-on at $z = 0$. The $z=0$ gas
distribution is plotted in the bottom-right panel of Fig.~\ref{z0}. }
\label{gasevolution}
\end{figure*}

\subsection{The evolution of angular momentum}\label{angmom}

Since the direction of the angular momentum vector of the main
progenitor halo varies in time due to gravitational torques, it is
expected that the direction of the angular momentum vector of the
accreting gas will also vary. Such a process may affect the formation
of the disc through angular momentum mixing.  In
Fig.~\ref{gasevolution}, we show the gas distribution around the main
progenitor in the shock-burst model. The viewing angle is fixed and
chosen so that the disc is edge-on at $z=0$ (see the bottom-right
panel of Fig.~\ref{z0}). A gas disc already exists at $z=1$, but its
orientation changes significantly throughout its evolution. For
example, the disc is inclined and almost edge-on at $z = 1$, but it
becomes almost face-on at $z = 0.7$ before settling down to the final
orientation at $z = 0.2$. A good example of angular momentum mixing is
seen in the panel for $z = 0.4$, where the newly accreting gas does
not line up with the pre-existing inner disc but settles instead onto
a different plane and the two discs torque one another. As these gas
discs host star formation, this process contributes to the formation
of hotter stellar components, such as a bulge or a thick disc.

\begin{figure}
\includegraphics[width=7.5cm]{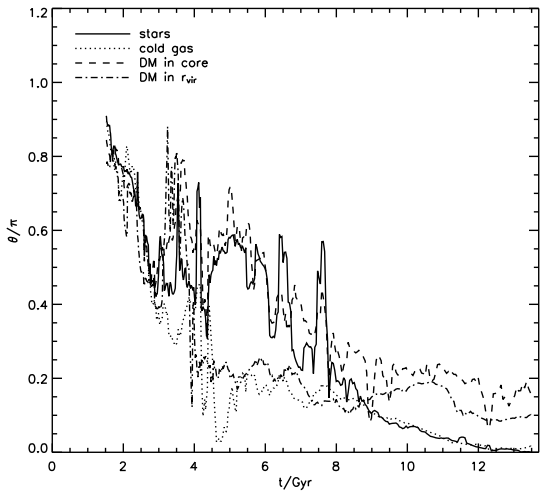}
\caption{
Evolution of the angles between the angular momentum vector of the
stellar component within $10 h^{-1}$ kpc at $z=0$ and the angular
momentum vectors of various components in the main progenitor for the
shock-burst simulation. The different lines refer to the following
components of the main progenitors: stars within $25 h^{-1}$ kpc
(solid), cold gas within $25 h^{-1}$ kpc (dotted), dark matter within
$25 h^{-1}$ kpc (dashed), and dark matter within the virial radius
(dot-dashed). Angles are divided by $\pi$, and $\theta/\pi=1$ means
counter-rotation relative to the final disc. }
\label{AM}
\end{figure}
In Fig.~\ref{AM}, we plot the evolution of the direction of the
angular momentum vectors of various components of the main progenitor
in the shock-burst model.  The lines show the angles between each
angular momentum vector and that of the final stellar disc. Angular
momenta are calculated for material inside a sphere of a radius $25
h^{-1}$ kpc (in physical coordinates) for stars (solid), cold gas
(dotted), and dark matter in the core (dashed), while the angular
momentum of the main progenitor halo is defined using the dark matter
inside the virial radius (dot-dashed). At early times, $t < 8.5$ Gyr,
the spin of the cold gas correlates well with that of the dark halo,
while the stars show a better correlation with the core at $t < 6$
Gyr. Note that the angular momentum of the stellar component is
contaminated by satellites and hence it is rather noisy. On the other
hand, at late times, $t > 8.5$ Gyr, corresponding to $z < 0.5$, the
stars follow the cold gas quite well and they are offset from both the
dark matter core and halo.

We confirmed that the angular momentum flip of the cold gas disc seen
in Figs.~\ref{gasevolution} is caused by minor mergers occurring
between $z \simeq 1$--$0.6$.  Afterwords, its evolution is driven by
angular momentum mixing with newly accreted gas that has a different
orientation from the pre-existing disc (see
Fig. \ref{gasevolution}). For the stellar component, initially the
direction of its angular momentum vector is well correlated with that
of the dark matter in the core, both of which result from merging and
dynamical friction.  As stars begin to form in the well-developed gas
disc, the nett stellar angular momentum becomes dominated by disc
stars, and the angular momentum vectors of the stellar and cold gas
component become aligned.

\subsection{Dependence on burst parameters}\label{dependence}

In this subsection, we discuss how galaxy properties are affected by
changes in the burst parameters. For the shock-burst model, we have
checked that if we adopt the same star formation efficiency in the
burst mode as in the quiescent mode, but retain the top-heavy IMF,
then the results are very similar to the no-burst simulation. Thus, a
top-heavy burst IMF alone is insufficient to produce a bigger
disc. Similarly, merely increasing the star formation efficiency in
the burst without including a top-heavy IMF does not lead to a
significantly bigger disc than is found in the no-burst simulation.
Only the combination of these two features, i.e. a top-heavy burst IMF
and an increased burst star formation efficiency ($c_* \sim 1$ in
eqn.~2) is capable of promoting stronger feedback to the extent that an
extended disc can form in this particular halo.

Since the response of the density-burst model to changes in the
density threshold is fairly straightforward and the shock-burst model
looks more promising, we decided to explore the following three
additional models:
\begin{itemize}
\item SH: The same as the shock-burst model, but with a slightly higher
threshold for the burst, $\dot{K}_{\rm burst} = 10^4$, rather than
$\dot{K}_{\rm burst} = 9 \times 10^3$ as in the standard shock-burst
model.
\item SL: The same as the shock-burst model, but with a lower threshold for 
the burst, $\dot{K}_{\rm burst} = 8 \times 10^3$. 
\item DS: A combination of density-induced and shock-induced burst
modes, with the same parameters as the original models.
\end{itemize}

\begin{figure}
\includegraphics[width=7.5cm]{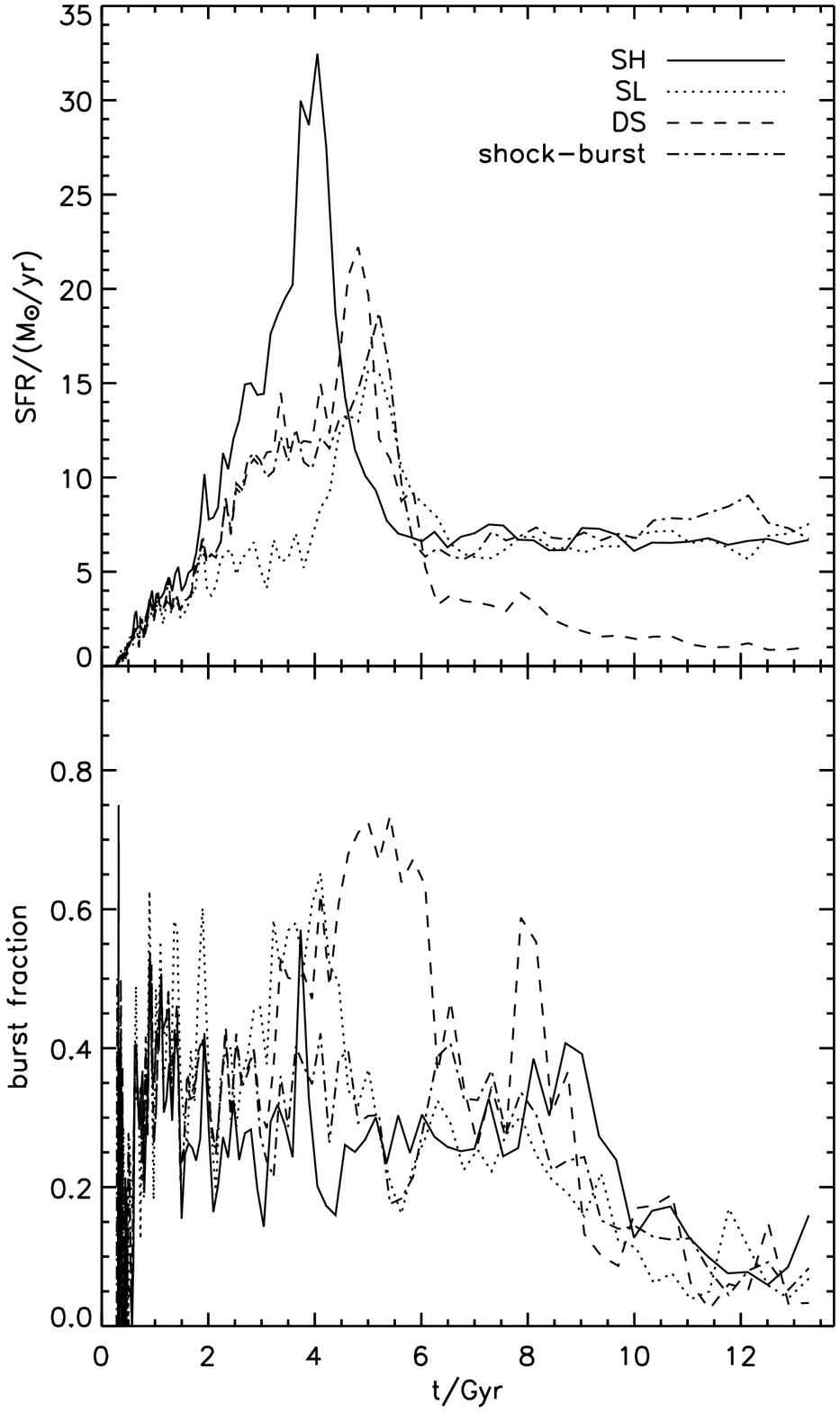}
\caption{
The same as Fig.~\ref{sfr} but for the SH (solid), SL (dotted), and DS
(dashed) models. The standard shock-density model is also shown for
comparison (dot-dashed). }
\label{sfr2}
\end{figure}
In Fig.~\ref{sfr2}, we show the SFHs of the galaxies in these three
variant models. Models with shock-induced bursts are quite sensitive
to the threshold, $\dot{K}_{\rm burst}$. In general, the lower the
threshold, the fewer the number of stars that form because of the
additional amount of feedback generated by the burst stars. The DS
model shows interesting behaviour. As the feedback from the
shock-induced bursts suppresses gas cooling at early times, the
redshift at which the gas density reaches the threshold for a
density-induced burst is reduced.  At this time, the depth of the
potential well is deeper and this decreases the strength of galactic
winds.  As a result, when the density-induced bursts start, the star
formation rate exceeds that in the shock-burst model. These
density-induced bursts in the DS model do not lead to a smaller final
bulge mass, but do suppress further cooling, showing how difficult it
is to promote disc formation when a large amount of feedback energy is
pumped into the galactic centre after $z \sim 2$.

\begin{figure*}
\includegraphics[height=11cm]{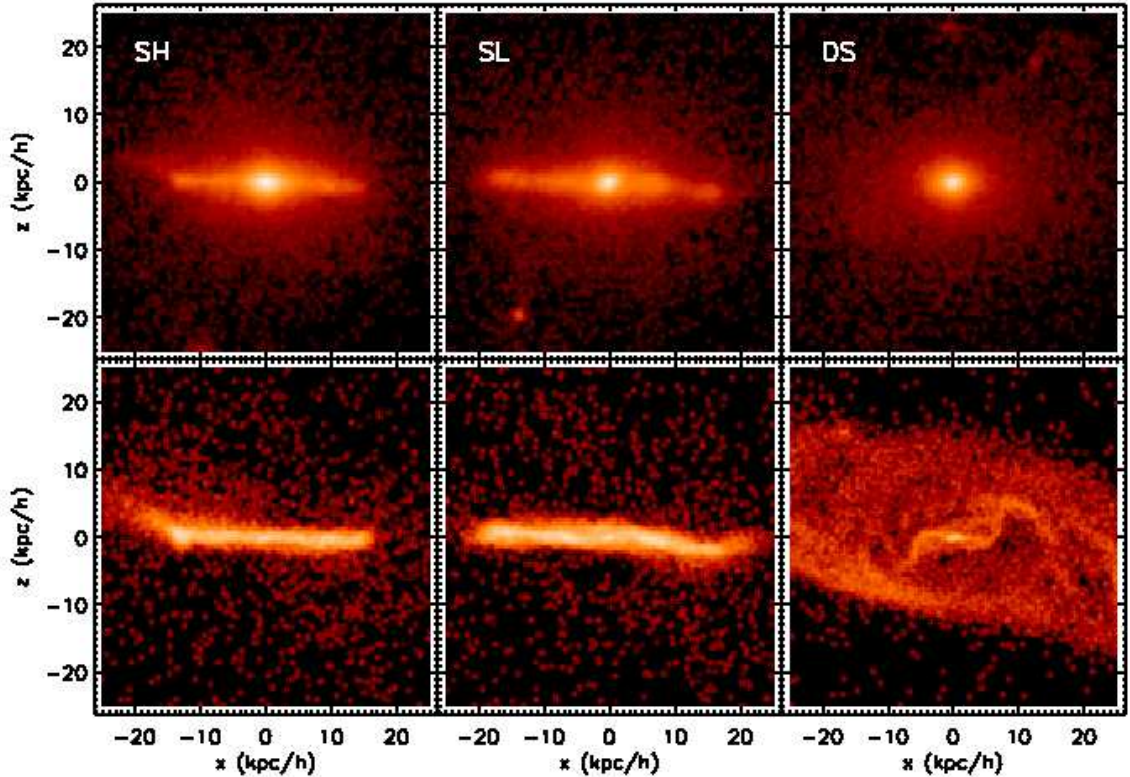}
\caption{
Edge-on projections of stars (upper row) and gas (lower row) in the SH,
SL, and DS models at $z=0$.}
\label{others}
\end{figure*}
Edge-on views of the final galaxies are shown in
Fig.~\ref{others}. Shock-induced bursts assist disc formation and
lowering the threshold for such bursts results in a larger disc.  Of
course, too low a threshold is counterproductive. For example, for
$\dot{K}_{\rm burst} = 5 \times 10^3$, the star formation becomes
dominated by the burst mode at all times. As a result, only a very dim
object forms because the threshold is below the typical values in the
accretion shock and most of the stars are then formed outside the
central disc. The galaxy in the DS model has an outer gas ring. One
might suspect that this is just gas swept out by feedback.  In fact,
the ring is newly accreted gas that had a different angular momentum
direction relative to that of the inner disc.  When the outer ring is
either spatially separated from the inner disc or perpendicular to it,
it tends to be stable. From our experience, such outer rings are not
rare in cosmological simulations because the direction of the angular
momentum of the main progenitor halo can change dramatically, as we
showed in Fig.~\ref{AM}, and so the orientation of newly accreting gas
can differ from that of the pre-existing disc
(see Fig.~\ref{gasevolution}).

\begin{table}
\caption{Disc-to-total ratios as in Table \ref{bt}, only
this time for galaxies in the SL, SH, and DS models.}
\label{bt2}
\begin{center}
\begin{tabular}{@{}lcccccc}
\hline
  & mass &
  $U$ &
  $B$ &
  $V$ & 
  $I$ & 
  $K$ \\
\hline 
SH & 0.39 & 0.83 & 0.80 & 0.76 & 0.66 & 0.59 \\
SL & 0.55 & 0.94 & 0.91 & 0.87 & 0.79 & 0.73 \\
DS & 0.20 & 0.51 & 0.47 & 0.41 & 0.33 & 0.28 \\
\hline
\end{tabular}
\end{center}
\end{table}
We also performed a dynamical decomposition for these galaxies (see Table
\ref{bt2}). As expected, lowering the threshold for a shock-induced
burst leads to larger disc-to-total ratios.  These numbers also
confirm that density-induced bursts result in early type galaxies even
when they are combined with shock-induced bursts.

\section{SUMMARY AND DISCUSSION}

We have performed cosmological simulations of galaxy formation in a
cold dark matter halo chosen to have a quiet merger history.  We have
employed a multi-phase description of the ISM based on that of
\citet{sh03} and developed explicit models for star formation in {\it quiescent} and
{\it burst} modes. We assume that the star formation efficiency is
much higher in the burst mode than in the quiescent mode and,
motivated by the semi-analytic work of \citet{bau04}, that the stars
in a burst form with a top-heavy IMF. Our model potentially provides a
unified picture of star formation in galaxies of all morphological
types. In this paper, we have explored two triggers for the starburst mode: 
strong shocks and a high density. An important conclusion of our
work is that, {\it for the same initial conditions}, different burst
models result in galaxies with a wide range of morphological types,
having bulge-to-total $B$-band luminosities spanning the range $D/T
\sim 0.2 $ to $0.9$.

Although, having simulated galaxy formation in only one halo, we must
be cautious about generalising our conclusions, the shock-induced
burst model does look promising for creating disc-dominated
galaxies in haloes with relatively quiet recent merger histories. The
lower the threshold for shock-induced bursts, the more the galaxy
becomes disc-dominated. The key to the success of this model in a
universe where structure grows hierarchically is that the burst
fraction is high at early times and subsequently decreases, once the
main phase of halo merging activity has taken place. Feedback from the
top-heavy IMF in the shock-induced bursts suppresses the early
collapse of baryons in small haloes, helping to create a reservoir of
hot gas that remains attached to the halo and is available for cooling
later on. It is this gas that eventually ends up forming a large,
young stellar disc in which stars continue to form at a slow rate up to
the present day. 

In spite of its success in generating a disc-dominated galaxy which
falls on the $I$-band Tully-Fisher relation, our shock-induced burst
model produced an excessively large disc with an unrealistically flat
surface density profile. A possible cause of this problem may be the
lack of collimation of the galactic winds in the simulation. The
pressure generated by the roughly isotropic winds causes the hot gas
reservoir to becomes more extended than the dark matter distribution
at early times and thus susceptible to gaining extra angular momentum
from tidal torques. When this gas later cools, it settles onto a disc
with too much angular momentum. Thus, far from having too little
angular momentum as most previously simulated discs, our disc ended up
with too much angular momentum! 
If this interpretation is correct, the
lack of collimation in the winds might just reflect the inability of
SPH to represent the steep pressure gradients required to collimate
the outflow. 
We intend to test this hypothesis with simulations using
a mesh-based hydrodynamic code.
This result cautions about the general assumption that the baryonic matter 
has the same specific angular momentum of the dark matter \citep{fe80}. 
In fact, feedback is able to change the distribution of the 
gas in proto-galactic regions, and therefore baryons can have higher 
angular momentum than dark matter. 

While other improvements implemented in our simulations, such as the
phase-decoupling method to suppress numerical transfer of angular
momentum, high resolution, and the use of a stiff equation of state
for the ISM, also contribute to the formation of large discs
\citep[see][] {oka03, gov04, rob04}, the comparison between models
with and without shock-induced bursts shows that our burst model is
the key ingredient that helps resolve the angular momentum problem in
a CDM universe.  Although the assumption of a top-heavy IMF in bursts
is controversial, it is encouraging that the model we have used is the
same that was originally introduced in order to account for the
abundance and properties of galaxy populations at high-redshift
\citep{bau04}, and was subsequently found also to provide a plausible explanation
for the metal content in the intracluster medium
\citep{nag04} and in elliptical galaxies \cite{nag05}. 

By contrast, the density-induced burst model seems to make things
worse for disc formation. Since the gas density only reaches the
threshold for bursts in massive haloes, this model does not affect gas
cooling and star formation in small haloes. The feedback associated
with the burst merely halts star formation in the galaxy at late times
when an extended stellar disc might otherwise be expected to form.
Consequently, the density-induced burst leads to the formation of an
early-type galaxy.  The effects of bursts (with their top-heavy IMF)
in this model may resemble those generated by energy fed back from
active galactic nuclei (AGN). While AGN do not produce stars and
metals, both forms of feedback inject large amounts of energy into the
galactic centre.  Thus, we suspect that it is difficult to solve the
angular momentum problem in simulations of disc galaxies solely by
including AGN feedback. Of course, it is entirely possible that AGN
feedback behaves quite differently from the case of density-induced
bursts (see \citet[]{kg04} and \citet{sdh04} for attempts at modelling
AGN feedback on galactic scales). However, if feedback from AGN does
have similar consequences to those produced by the density bursts in
our simulations, then one might expect that it is particularly
important for suppressing recent star formation in the most massive
galaxies. This is in qualitative agreement with the observation that
the most massive galaxies tend to be ellipsoidal and not discy.

The morphologies of galaxies formed in the simulations with shock-induced 
bursts are rather sensitive to the value of the burst threshold 
$\dot{K}_{\rm burst}$, as shown in \S4.4, because the allowed range for 
this parameter is very narrow. 
With too low a threshold, the star formation is dominated 
by the burst mode at all times and vice versa. As a result, the allowed 
range for $\dot{K}_{\rm burst}$ is less than an order of magnitude in 
our simulations. 
Within this range, the behaviour of the shock-burst models is easily 
understood. Lower threshold values suppress star formation more effectively 
at high redshift. Interestingly, the three shock-burst models have almost 
the same star formation rate at low redshift, when no major mergers occur. 
Since our current modelling of shock-induced bursts is not resolution 
independent as we discussed in \S2.2.4, we will explore alternative
formulations in future work. 

Several important tests of our model are possible and we intend to
pursue these in future work. For example, we have followed in detail
the evolution of the metallicity of gas and stars in our simulations,
keeping track of metals produced in the quiescent and burst
modes. Comparing the resulting distributions of heavy elements in the
different dynamical components of the simulated galaxies with
observational data will be an important test of the validity of our
feedback models.  A key question is how often do shock-induced bursts
like those we have modelled lead to the formation of disc-dominated
galaxies. Answering this question will require simulating galaxy
formation in large volumes. The results presented in this paper
encourage us to believe that such demanding calculations are worth
pursuing. 

\section*{Acknowledgments}
We are grateful to Volker Springel for providing us with the GADGET2
code. We also thank Masahiro Nagashima who made look-up tables for
chemical evolution. The simulations were performed on the Cosmology
Machine at the Institute for Computational Cosmology in the University
of Durham. This work was supported by a PPARC rolling grant for
Extragalactic Astronomy and Cosmology. TO is a Research Fellow of the
Japan Society for the Promotion of Science (No.01891). VRE
acknowledges a Royal Society University Research Fellowship.

\bsp

\label{lastpage}

\end{document}